\documentclass[aps,prd,amsfonts, eqsecnum,nofootinbib,twocolumn]
{revtex4} 
\usepackage{graphics,epsfig}
\begin{document}
\preprint{MSU-041021}
%
% Title of paper
\title{Electroweak Symmetry Breaking: \\
{\sl With Dynamics}\footnote{Talks presented at IXth Mexican Workshop
on Particles and Fields, Colima, Mexico, Nov. 17-22, 2003 and the 15th Topical
Conference on Hadron Collider Physics, HCP2004, Michigan
State University, June 14-18, 2004.}}
 \vskip 0.1in
\author{R. Sekhar Chivukula} \email[]{sekhar@msu.edu}
\affiliation{Department of Physics and Astronomy\\
  Michigan State University \\
  East Lansing, MI 48824, USA}
\date{October 2004}
\begin{abstract}
In this note I provide a brief description of models of dynamical
electroweak symmetry breaking, including walking technicolor, top-color
assisted technicolor, the top-quark seesaw model, and little higgs theories.
\end{abstract}
\pacs{}
\maketitle

%%%%%%%%%%%%%%%%%%%%%%%%%%%%%%%%%%%%%%%%%%%%%%%%%%%%%%%%%%%%%%%%%%%%

%%%%%%%%%%%%%%%%%%%%%%%%%%%%%%%%%%%%%%%%%%%%
%% MAINMATTER
%%%%%%%%%%%%%%%%%%%%%%%%%%%%%%%%%%%%%%%%%%%%

\section{Shortcomings of the Fundamental Higgs Model}

The standard electroweak theory is in spectacular agreement with precision
electroweak data \cite{unknown:2003ih}. Unfortunately, the theory is manifestly incomplete:
while we know that the electroweak gauge symmetry must be spontaneously broken
down to electromagnetism, we have no direct experimental evidence for the agent responsible
for this symmetry breaking. The simplest choice, the one-Higgs doublet model, has a number
of shortcomings, namely:

\begin{enumerate}

\item It is a theory of a fundamental scalar particle and we have, so far, observed 
no fundamental scalars in nature.

\item The model provides no explanation for why electroweak symmetry
breaking occurs, or why it occurs at the weak scale.

\item It suffers from the hierarchy and naturalness problems -- namely, if there is
physics at some much higher energy (say the GUT or Planck scale), then a
precise adjustment of the underlying parameters (i.e. fine-tuning) is required
to keep the electroweak scale low.

\item The model is trivial: it does not exist as a continuum
quantum field theory.

\end{enumerate}

Given our absence of understanding about the nature of electroweak symmetry
breaking, it is important to remember why the Higgs or so other 
additional dynamics is necessary. What
is wrong with simply adding masses to the gauge bosons of the electroweak theory?
Any scattering amplitude in a consistent quantum mechanical theory must lie within
the unitarity circle. However, if we compute the scattering amplitude of longitudinal
W bosons at tree-level, using the interactions of SU(2) x U(1) gauge theory, at
high-energies we find contributions which grow with energy
\cite{Dicus:1992vj,Cornwall:1973tb,Cornwall:1974km,Lee:1977yc,Lee:1977eg,Veltman:1976rt}. 
The leading contributions,
which grow like energy to the fourth power, cancel due to the symmetries of the
underlying nonabelian gauge theory. Unfortunately, the same cannot be
said of the subleading, energy-squared, divergence.

In the absence of additional physics contributing to the amplitude, we see
that tree-level unitarity is violated at an energy of order 1 TeV -- hence we conclude
that some new physics is required below this energy scale. The Higgs boson, the
remnant scalar degree of freedom in the one-Higgs doublet model, has couplings
precisely adjusted (again, as a result of the underlying gauge symmetry) so as to
precisely cancel this bad high-energy behavior. In this talk, we will focus on theories
in which electroweak symmetry breaking occurs due to new strong 
dynamics at energy scales of order 1 TeV.\footnote{For a recent comprehensive review
of these models, and a complete set of references, see \protect\cite{Hill:2002ap}.}

\section{Technicolor: Higgsless since 1976!}

So, if not a Higgs boson, what might exist at energies of order 1 TeV or
less and be responsible for electroweak symmetry breaking? The most elegant
possibility is that electroweak symmetry breaking arises from chiral symmetry
breaking due to a new, strongly-interacting, gauge theory: Technicolor
\cite{Weinberg:1975gm,Susskind:1978ms}. 
In the simplest such model one introduces a new strong $SU(N_{TC})$
gauge theory and, analogous to the up- and down-quarks in QCD, 
two new fermions transforming (which we will
denote $U$ and $D$) as fundamentals
of this gauge symmetry. These new ``techniquarks'' carry an $SU(2)_L \times SU(2)_R$
global symmetry -- the analog of the (approximate) chiral symmetry of the light 
quarks in QCD. Just as in QCD, the ``low-energy'' strong dynamics of this
new gauge theory is expected to cause chiral symmetry breaking, that is a non-perturbative
expectation value for the chiral condensates $\langle \bar{U}_L U_R \rangle =
\langle \bar{U}_R U_L\rangle$ and similarly for the $D$ fermions.

If the left-handed techniquarks form an $SU(2)_W$ doublet, while the right-handed
techniquarks are weak singlets carrying hypercharge, 
technicolor chiral symmetry breaking will result in electroweak 
symmetry breaking. The Goldstone bosons arising from chiral symmetry breaking are
transmuted, by the Higgs mechanism, into the longitudinal components of the electroweak
gauge bosons. 

Theoretically, technicolor addresses all of the shortcomings of the one-doublet
Higgs model: there are no scalars, electroweak symmetry breaking arises in a 
natural manner due to the strong dynamics of a non-abelian gauge theory, 
the weak scale is related to the renormalization group flow of the strong technicolor coupling -- 
and can be much smaller than any high energy scale and, due to asymptotic freedom, the
theory (most likely) exists in a rigorous sense.

Unfortunately, the simplest versions of this theory -- based, as described, on a scaled-up
version of QCD -- are not compatible with precision electroweak data\footnote{See Langacker and
Erler in \protect\cite{Eidelman:2004wy}.} (and, as described
so far, cannot accommodate the masses of the quarks and leptons). Nonetheless, this simplest
version remains a paradigm for thinking about theories of dynamical electroweak symmetry
breaking.

\begin{figure}
  \includegraphics[height=.2\textheight]{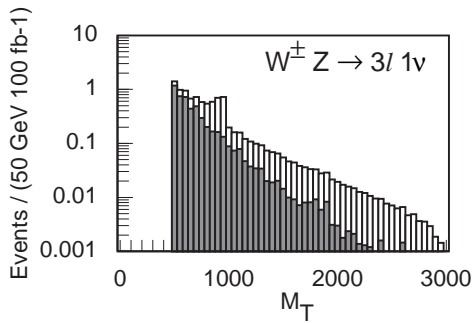}
  \caption{Event rate for $W^\pm Z$ scattering in ``gold-plated'' mode at the LHC, for a 1 TeV
  technirho meson \protect\cite{Golden:1995xv}.}
\end{figure}

In the final analysis, the agent of electroweak symmetry breaking will be uncovered
by experiment -- not by theoretical investigation or prejudice. Although a scaled-up version
of QCD is not a viable model of electroweak symmetry breaking,  we may use our 
knowledge of QCD to investigate the possible collider signatures of such a sector
at the Tevatron or LHC. In QCD, the most prominent resonances in pion scattering
are the vector mesons. Analogously, in a theory of dynamical electroweak symmetry
breaking, we expect the most prominent resonances in longitudinal $W$ scattering will
be ``technivector'' mesons. An illustration of the event rates in a minimal model at the LHC
are shown in figure 1. This minimal model is not particularly encouraging -- the number of
events at the LHC at high luminosity is barely enough to see a 1 TeV resonance, much
less a heavier one! Fortunately, as we will see, this minimal model is likely to severely
understate the range and accessibility of signatures of a model of dynamical electroweak 
symmetry breaking in a realistic model which can accommodate quark and lepton masses.

\section{Fermion Masses and ETC Interactions}

While technicolor provides a natural paradigm for producing masses for
the electroweak gauge bosons, the agent of electroweak symmetry breaking must
also provide for fermion masses. In the fundamental higgs model, one introduces yukawa
interactions between the fermions and the higgs with couplings proportional to the desired
fermion masses. In a model of dynamical electroweak symmetry breaking, one must couple
fermions to the electroweak symmetry-breaking condensate. A natural model for doing so involves
enlarging the technicolor group and embedding (some or all) quark flavor into the group
as shown below. 

These models, known as ``extended technicolor models'' (ETC)
\cite{Eichten:1979ah,Dimopoulos:1979es}, relate the fermion masses
to the masses of the extended technicolor gauge bosons responsible for coupling
the fermions to the symmetry breaking condensate. In the case of a QCD-like theory, 
in which the techniclor theory is precociously asymptotically free, we may
estimate the scale of masses of the ETC gauge bosons to be
\begin{equation}
{M_{ETC}\over g_{ETC}}\simeq 40\,{\rm TeV}\,
\left({{F_{TC}} \over 250\,{\rm GeV}}\right)^{3\over 2}
\left({100\,{\rm MeV} \over m_q} \right)^{1\over 2}~.
\end{equation}

Extended technicolor is a tremendously ambitious theory: in addition to electroweak symmetry
breaking, it recasts the problem of quark and lepton flavor in terms of the breaking of the extended
technicolor interactions. In principle, this is progress -- one could conceivably construct a dynamical
explanation of flavor! In practice, however, there is a substantial obstacle \cite{Eichten:1979ah}. 
In order to give rise
to the various different fermion masses and mixings, the ETC interactions must distinguish
amongst the various flavors of quarks (and leptons) -- e.g. distinguish strange-quarks from down-
or bottom-quarks. In general, such interactions will give rise to flavor-changing neutral
currents. The limits on flavor-changing interactions of the strange-quark, for example,
imply that the scale associated with such interactions must exceed 500-1000 TeV.\footnote{For
a recent comprehensive review of flavor constraints on ETC theories, see \protect\cite{Appelquist:2004ai}.}

Extended technicolor gauge bosons of that mass, however, can (at least in QCD-like
technicolor) only accommodate a quark mass of order 1-10 MeV -- much too small for
the strange- or charm-quark masses, much less the third generation.
From this, along with constraints on precision electroweak parameters, we conclude that
technicolor dynamics cannot be QCD-like.

\section{Walking Technicolor and Beyond}

A proposal for how technicolor could differ from QCD goes under the name
of ``walking'' technicolor. It has been argued \cite{Holdom:1981rm,Holdom:1984sk,Yamawaki:1985zg,Appelquist:1986an,Appelquist:1986tr,Appelquist:1987fc}
that if the $\beta$-function
for the technicolor coupling, which controls how quickly the coupling constant falls
as one scales to higher energy, is small, than the anomalous dimension for the technifermion
mass operator (the function $\gamma_m$ above) will be close to one for a large range
of energies. Such a change enhances the quark or lepton masses allowed, perhaps
as high as a few GeV -- enough, perhaps, for the first and second generations of quarks 
and leptons.

In order to have a small $\beta$-function, the technicolor theory must
have many fermions, perhaps in several representations of the technicolor group.
Generically, this implies that there can be several different scales of technicolor
chiral symmetry breaking \cite{Eichten:1996dx,Eichten:1997yq}. 
The phenomenology of these ``multiscale'' models
could involve a large approximate chiral symmetry group, and therefore many
light ``pseudo-Goldstone'' bosons, and potentially light technivector mesons.
These considerations have driven the searches for technicolor at LEP
and the Tevatron. \footnote{For a review, see Chivukula, Narain, and Womersley
in \protect\cite{Eidelman:2004wy}.} It should be emphasized that, so far, these
searches are rather model dependent and are just reaching the interesting
regime. Run II of the Tevatron and the LHC will substantially extend these
limits.

\section{The Top Quark}

While walking technicolor may be sufficient to produces masses for the
first two generations, it is unlikely to be able to do so for the third generation
and for the top-quark in particular.\footnote{For a valiant attempt to produce 
all of the quark masses and mixings in a walking technicolor theory, see 
\protect\cite{Appelquist:2003hn}.} The
difficulty with the top-quark is easy to see, top-quark
mass generation in an ETC theory implies that
\begin{equation}
{M_{ETC} \over g_{ETC}} \simeq 1\,{\rm TeV}\,
\left({{F_{TC}} \over 250\,{\rm GeV}}\right)^{3\over 2}
\left({175\,{\rm GeV} \over m_t} \right)^{1\over 2}~,
\end{equation}
where $M_{ETC}$ and $g_{ETC}$ are the masses and couplings of the ETC
gauge-bosons responsible for top-quark mass generation, and $F_{TC}$ is
the technicolor ``$F$''-constant (analogous to $f_\pi$ in QCD) which cannot
be higher than 250 GeV.
We see that, given the mass of the top-quark, the ETC gauge bosons 
required are very light -- so light, that there
is little distinction between them and the technicolor interactions, leading to
potential problems with $Z\to \bar{b} b$ \cite{Chivukula:1992ap} and 
$\Delta \rho$ \cite{Chivukula:1988qr}.

These considerations suggest that there may be a separate sector associated
with generation of the top quark mass. Topcolor Assisted technicolor
\cite{Hill:1994hp} is
such a theory. In this model, technicolor is responsible for the bulk of electroweak symmetry
breaking, and extended technicolor for the masses of the light quarks and leptons. An
additional strong color sector, coupling only to the third generation, generates
a nonzero condensate of top quarks ($\langle \bar{t} t \rangle \neq 0$), and gives rise
to a large topquark mass. The simplest scheme incorporates 
two color groups, the stronger of which couples
only to the third generation and breaks down to ordinary color at a scale of order
1 TeV, leaving a color octet of ``topgluons.''  An additional copy of hypercharge 
distinguishes the top-quark from the bottom-quark, leaving a heavy $Z'$ boson
with flavor-dependent couplings. The topgluons are a particularly novel
phenomenological feature of these models, and illustrate the possibility of interesting signals
involving $b$- and $t$-quark jets \cite{Harris:1999ya}.

\section{Composite Higgs Bosons: Top Seesaw and Little Higgs}

If topcolor with technicolor is good, perhaps we don't need technicolor! That is,
perhaps the top quark plays the role of a technifermion, and a top quark condensate
is responsible for all of electroweak symmetry breaking 
\cite{Miransky:1989ds,Miransky:1988xi,Marciano:1989xd,Nambu,Bardeen:1989ds}. At first
sight, this would seem difficult: the top quark mass is of order 175 GeV, while the
value of the electroweak scale (the expectation value of the Higgs field in the standard
scalar doublet model) is 250 GeV. Our intuition from QCD, bolstered by model calculations,
is that if a top-quark condensate is responsible for electroweak symmetry breaking, then
the top-quark should have a mass of order 500 GeV. 

Dobrescu and Hill have constructed
an elegant alternative \cite{Dobrescu:1997nm}.  Namely,
they propose that there is a new $SU(2)_W$-singlet set of quarks
$\chi_{L,R}$, and the electroweak symmetry breaking condensate is of the form
$\langle \bar{\chi}_R t_L \rangle$. If this were the whole story, the only non-zero mass
quark would have a mass of around 500 GeV, as discussed above. 
However, Dobrescu and Hill propose that this
state mixes via a seesaw-type mass matrix, and that the lowest mass eigenstate of the
full $\chi - t$ system is to be identified with the top-quark.

Interestingly, the same dynamics that produces an electroweak symmetry breaking
condensate in the top seesaw model produces a scalar bound state that couples
approximately like a Higgs boson \cite{Chivukula:1998wd,He:2001fz}.
Light scalar bosons without supersymmetry
typically require fine-tuning, and the top seesaw model is no exception. The top condensate
is, we presume, driven by strong, short-distance topcolor interactions with a natural scale
of order a TeV or higher. The strength of these interactions must be adjusted carefully to produce an
effective composite Higgs boson vacuum expectation value of order only a few hundred
GeV -- this adjustment is a dynamical manifestation of the fine-tuning that we expect
in a scalar theory. Here, however, the underlying scale is only of order a TeV or so, and the
fine-tuning is not nearly as severe as would be required in a scalar GUT theory.

Recently, a new class of models with a naturally light composite Higgs boson
has emerged: ``little higgs'' theories \cite{Arkani-Hamed:2001nc,Arkani-Hamed:2002pa,Arkani-Hamed:2002qy,Arkani-Hamed:2002qx}. In these models the Higgs boson is one of many
pseudo-Goldstone bosons whose mass is protected by a spontaneously broken
chiral symmetry -- in analogy to why the pion mass remains light in QCD. Inspired
by investigations of ``deconstructed'' higher-dimensional gauge theories
\cite{Arkani-Hamed:2001ca,Hill:2000mu}, the chiral
symmetries of the models are constrained in such a way that a Higgs-boson self-coupling
can appear without the appearance of large corrections to the Higgs boson mass. In general
the large contributions to the Higgs-boson mass arising from top-quark
loops are cancelled by corrections from a new singlet quark (similar to the $\chi$
introduced in top seesaw models) and contributions arising from electroweak gauge bosons
are cancelled by those from an extended electroweak symmetry. Unlike supersymmetry,
the cancellation occurs between contributions from particles {\it of the same spin} --
with the cancellation enforced by the underlying chiral symmetries. The properties
of these models below an energy scale of 10 TeV or so depend only on the symmetries of
the model and are independent of the underlying dynamics. Such a scenario suggests that
the LHC would uncover only the beginning of a rich new set of dynamics, and that a very
high-energy hadron collider would be required to examine the underlying theory.

\section{Conclusions}

\begin{itemize}

\item A fundamental standard model Higgs is {\bf unnatural}, {\bf unattractive}, and 
(so far) {\bf unobserved}.

\item Strong Dynamics, such as technicolor,  provides an elegant dynamical explanation for 
EWSB, but is challenged by precision electroweak tests and FCNCs. These considerations
drive the investigation of composite higgs models of various kinds.

\item Current limits (from Tevatron and LEP) are just reaching the interesting regime to test
models of dynamical electroweak symmetry breaking. There are important signatures 
of these models involving W- and  Z-bosons, and t- and b-quarks.

\item Strong Dynamics associated with electroweak symmetry breaking will be 
discovered (or ruled out) by experiments at hadron colliders in this decade

\end{itemize}

%%%%%%%%%%%%%%%%%%%%%%%%%%%%%%%%%%%%%%%%%%%%%%%%
%% BACKMATTER
%%%%%%%%%%%%%%%%%%%%%%%%%%%%%%%%%%%%%%%%%%%%%%%%

%\begin{theacknowledgments}
 
%\end{theacknowledgments}

%% If the bibliography is produced without BibTeX comment out the


%\bibliography{hcp2004}

\begin{thebibliography}{38}
\expandafter\ifx\csname natexlab\endcsname\relax\def\natexlab#1{#1}\fi
\providecommand{\enquote}[1]{``#1''}
\expandafter\ifx\csname url\endcsname\relax
  \def\url#1{\texttt{#1}}\fi
\expandafter\ifx\csname urlprefix\endcsname\relax\def\urlprefix{URL }\fi

\bibitem[unk(2003)]{unknown:2003ih}LEP Electroweak Working Group,
{\tt http://lepewwg.web.cern.ch/LEPEWWG/} .

\bibitem[Dicus and Mathur(1973)]{Dicus:1992vj}
Dicus, D.~A., and Mathur, V.~S., \emph{Phys. Rev.}, \textbf{D7}, 3111--3114
  (1973).

\bibitem[Cornwall et~al.(1973)]{Cornwall:1973tb}
Cornwall, J.~M., Levin, D.~N., and Tiktopoulos, G., \emph{Phys. Rev. Lett.},
  \textbf{30}, 1268--1270 (1973).

\bibitem[Cornwall et~al.(1974)]{Cornwall:1974km}
Cornwall, J.~M., Levin, D.~N., and Tiktopoulos, G., \emph{Phys. Rev.},
  \textbf{D10}, 1145 (1974).

\bibitem[Lee et~al.(1977{\natexlab{a}})]{Lee:1977yc}
Lee, B.~W., Quigg, C., and Thacker, H.~B., \emph{Phys. Rev. Lett.},
  \textbf{38}, 883 (1977{\natexlab{a}}).

\bibitem[Lee et~al.(1977{\natexlab{b}})]{Lee:1977eg}
Lee, B.~W., Quigg, C., and Thacker, H.~B., \emph{Phys. Rev.}, \textbf{D16},
  1519 (1977{\natexlab{b}}).

\bibitem[Veltman(1977)]{Veltman:1976rt}
Veltman, M. J.~G., \emph{Acta Phys. Polon.}, \textbf{B8}, 475 (1977).

\bibitem[Hill and Simmons(2003)]{Hill:2002ap}
Hill, C.~T., and Simmons, E.~H., \emph{Phys. Rept.}, \textbf{381}, 235--402
  (2003).

\bibitem[Weinberg(1976)]{Weinberg:1975gm}
Weinberg, S., \emph{Phys. Rev.}, \textbf{D13}, 974--996 (1976).

\bibitem[Susskind(1979)]{Susskind:1978ms}
Susskind, L., \emph{Phys. Rev.}, \textbf{D20}, 2619--2625 (1979).

\bibitem[Eidelman et~al.(2004)]{Eidelman:2004wy}
Eidelman, S., et~al., \emph{Phys. Lett.}, \textbf{B592}, 1 (2004).

\bibitem[Golden et~al.(1995)]{Golden:1995xv}
Golden, M., Han, T., and Valencia, G. hep-ph/9511206 (1995).

\bibitem[Eichten and Lane(1980)]{Eichten:1979ah}
Eichten, E., and Lane, K.~D., \emph{Phys. Lett.}, \textbf{B90}, 125--130
  (1980).

\bibitem[Dimopoulos and Susskind(1979)]{Dimopoulos:1979es}
Dimopoulos, S., and Susskind, L., \emph{Nucl. Phys.}, \textbf{B155}, 237--252
  (1979).

\bibitem[Appelquist et~al.(2004{\natexlab{a}})]{Appelquist:2004ai}
Appelquist, T., Christensen, N., Piai, M., and Shrock, R. (2004{\natexlab{a}}).

\bibitem[Holdom(1981)]{Holdom:1981rm}
Holdom, B., \emph{Phys. Rev.}, \textbf{D24}, 1441 (1981).

\bibitem[Holdom(1985)]{Holdom:1984sk}
Holdom, B., \emph{Phys. Lett.}, \textbf{B150}, 301 (1985).

\bibitem[Yamawaki et~al.(1986)]{Yamawaki:1985zg}
Yamawaki, K., Bando, M., and Matumoto, K.-i., \emph{Phys. Rev. Lett.},
  \textbf{56}, 1335 (1986).

\bibitem[Appelquist et~al.(1986)]{Appelquist:1986an}
Appelquist, T.~W., Karabali, D., and Wijewardhana, L. C.~R., \emph{Phys. Rev.
  Lett.}, \textbf{57}, 957 (1986).

\bibitem[Appelquist and Wijewardhana(1987{\natexlab{a}})]{Appelquist:1986tr}
Appelquist, T., and Wijewardhana, L. C.~R., \emph{Phys. Rev.}, \textbf{D35},
  774 (1987{\natexlab{a}}).

\bibitem[Appelquist and Wijewardhana(1987{\natexlab{b}})]{Appelquist:1987fc}
Appelquist, T., and Wijewardhana, L. C.~R., \emph{Phys. Rev.}, \textbf{D36},
  568 (1987{\natexlab{b}}).

\bibitem[Eichten and Lane(1996)]{Eichten:1996dx}
Eichten, E., and Lane, K.~D., \emph{Phys. Lett.}, \textbf{B388}, 803--807
  (1996).

\bibitem[Eichten et~al.(1997)]{Eichten:1997yq}
Eichten, E., Lane, K.~D., and Womersley, J., \emph{Phys. Lett.}, \textbf{B405},
  305--311 (1997).

\bibitem[Appelquist et~al.(2004{\natexlab{b}})]{Appelquist:2003hn}
Appelquist, T., Piai, M., and Shrock, R., \emph{Phys. Rev.}, \textbf{D69},
  015002 (2004{\natexlab{b}}).

\bibitem[Chivukula et~al.(1992)]{Chivukula:1992ap}
Chivukula, R.~S., Selipsky, S.~B., and Simmons, E.~H., \emph{Phys. Rev. Lett.},
  \textbf{69}, 575--577 (1992).

\bibitem[Chivukula(1988)]{Chivukula:1988qr}
Chivukula, R.~S., \emph{Phys. Rev. Lett.}, \textbf{61}, 2657 (1988).

\bibitem[Hill(1995)]{Hill:1994hp}
Hill, C.~T., \emph{Phys. Lett.}, \textbf{B345}, 483--489 (1995).

\bibitem[Harris et~al.(1999)]{Harris:1999ya}
Harris, R.~M., Hill, C.~T., and Parke, S.~J., hep-ph/9911288  (1999).

%\cite{Miransky:1989ds}
\bibitem{Miransky:1989ds}
V.~A.~Miransky, M.~Tanabashi and K.~Yamawaki,
%``Is The T Quark Responsible For The Mass Of W And Z Bosons?,''
Mod.\ Phys.\ Lett.\ A {\bf 4}, 1043 (1989).
%%CITATION = MPLAE,A4,1043;%%

%\cite{Miransky:1988xi}
\bibitem{Miransky:1988xi}
V.~A.~Miransky, M.~Tanabashi and K.~Yamawaki,
%``Dynamical Electroweak Symmetry Breaking With Large Anomalous Dimension And T
%Quark Condensate,''
Phys.\ Lett.\ B {\bf 221}, 177 (1989).
%%CITATION = PHLTA,B221,177;%%

%\cite{Marciano:1989xd}
\bibitem{Marciano:1989xd}
W.~J.~Marciano,
%``Heavy Top Quark Mass Predictions,''
Phys.\ Rev.\ Lett.\  {\bf 62}, 2793 (1989).
%%CITATION = PRLTA,62,2793;%%

%\cite{Nambu}
\bibitem{Nambu}
Y. Nambu, EFI-89-08 (1989).

\bibitem[Bardeen et~al.(1990)]{Bardeen:1989ds}
Bardeen, W.~A., Hill, C.~T., and Lindner, M., \emph{Phys. Rev.}, \textbf{D41},
  1647 (1990).

\bibitem[Dobrescu and Hill(1998)]{Dobrescu:1997nm}
Dobrescu, B.~A., and Hill, C.~T., \emph{Phys. Rev. Lett.}, \textbf{81},
  2634--2637 (1998).

\bibitem[Chivukula et~al.(1999)]{Chivukula:1998wd}
Chivukula, R.~S., Dobrescu, B.~A., Georgi, H., and Hill, C.~T., \emph{Phys.
  Rev.}, \textbf{D59}, 075003 (1999).

\bibitem[He et~al.(2002)]{He:2001fz}
He, H.-J., Hill, C.~T., and Tait, T. M.~P., \emph{Phys. Rev.}, \textbf{D65},
  055006 (2002).

\bibitem[Arkani-Hamed et~al.(2001{\natexlab{a}})]{Arkani-Hamed:2001nc}
Arkani-Hamed, N., Cohen, A.~G., and Georgi, H., \emph{Phys. Lett.},
  \textbf{B513}, 232--240 (2001{\natexlab{a}}).

\bibitem[Arkani-Hamed et~al.(2002{\natexlab{a}})]{Arkani-Hamed:2002pa}
Arkani-Hamed, N., Cohen, A.~G., Gregoire, T., and Wacker, J.~G., \emph{JHEP},
  \textbf{08}, 020 (2002{\natexlab{a}}).

\bibitem[Arkani-Hamed et~al.(2002{\natexlab{b}})]{Arkani-Hamed:2002qy}
Arkani-Hamed, N., Cohen, A.~G., Katz, E., and Nelson, A.~E., \emph{JHEP},
  \textbf{07}, 034 (2002{\natexlab{b}}).

\bibitem[Arkani-Hamed et~al.(2002{\natexlab{c}})]{Arkani-Hamed:2002qx}
Arkani-Hamed, N., et~al., \emph{JHEP}, \textbf{08}, 021 (2002{\natexlab{c}}).

\bibitem[Arkani-Hamed et~al.(2001{\natexlab{b}})]{Arkani-Hamed:2001ca}
Arkani-Hamed, N., Cohen, A.~G., and Georgi, H., \emph{Phys. Rev. Lett.},
  \textbf{86}, 4757--4761 (2001{\natexlab{b}}).

\bibitem[Hill et~al.(2001)]{Hill:2000mu}
Hill, C.~T., Pokorski, S., and Wang, J., \emph{Phys. Rev.}, \textbf{D64},
  105005 (2001).

\end{thebibliography}

% \typeout{******************************************}
% \typeout{** Please run "bibtex \jobname" to optain}
%  \typeout{** the bibliography and then re-run LaTeX}
% \typeout{** twice to fix the references!}
% \typeout{******************************************}
%\typeout{}
% }

\end{document}